# Far-ultraviolet signatures of the $^{3}$He($n,tp$) reaction in noble gas mixtures

Patrick P. Hughes,[1)] Michael A. Coplan,[2] Alan K. Thompson,[1] Robert E. Vest[1] and Charles W. Clark[1,2,3]

[1]*National Institute of Standards and Technology, Gaithersburg, MD 20899*
[2]*Institute for Physical Science and Technology, University of Maryland, College Park, MD 20742*
[3]*Joint Quantum Institute, National Institute of Standards and Technology and University of Maryland, Gaithersburg, MD 20899*

**Abstract:** Previous work showed that the $^{3}$He($n,tp$) reaction in a cell of $^{3}$He at atmospheric pressure generated tens of far-ultraviolet photons per reacted neutron. Here we report amplification of that signal by factors of 1000 and more when noble gases are added to the cell. Calibrated filter-detector measurements show that this large signal is due to noble-gas excimer emissions, and that the nuclear reaction energy is converted to far-ultraviolet radiation with efficiencies of up to 30%. The results have been placed on an absolute scale through calibrations at the NIST SURF III synchrotron. They suggest possibilities for high-efficiency neutron detectors as an alternative to existing proportional counters.

The $^{3}$He($n,tp$) process, in which a neutron reacts with a helion to produce a proton and a triton with excess energy of 764 keV, is one of the best-characterized neutron reactions [1]. This reaction is the trigger mechanism for the $^{3}$He proportional counter, which is presently one of the most widely-used neutron detectors [2]. Here we discuss using the same trigger reaction to initiate far-ultraviolet (FUV) optical emissions, rather than electrical discharges. We have found cases in which about one-third of the nuclear reaction energy is converted to FUV emissions by noble-gas excimer molecules. Such high conversion efficiencies have been encountered previously in electric discharges [3-6], irradiation of noble gases by beams of electrons [7-9] and proton [10] bombardment of noble gases. Here they are reported for the first time in the context of a nuclear reaction and put on a quantitative basis needed to evaluate their applicability in a new FUV-based neutron detector.

Following previous work [11], we started with a gas cell containing pure $^{3}$He and $^{3}$He/$^{4}$He mixtures, exposed to neutrons with a deBroglie wavelength of 0.384 ± 0.004 nm, produced at the NIST Center for Neutron Research. At this wavelength, the cross section for the reaction $^{3}$He($n,tp$) reaction is (1.135 ± 0.012) x $10^{-20}$ cm$^{2}$. We measured FUV emissions in the cell with a filter-detector package that was calibrated at the NIST SURF III Synchrotron Ultraviolet Radiation Facility. This package is discussed below, as are the procedures for absolute measurements of the neutron flux and modeling of the optical detection system. The combination of measurement and modeling enables us to determine the number of FUV photons produced per neutron reaction.

In the work with pure $^{3}$He and $^{3}$He/$^{4}$He mixtures the source of the optical signal was determined to be Lyman α radiation from the 2p states of the $^{1}$H and $^{3}$H formed as neutralization and excitation products of the $^{3}$He($n,tp$) process [11,12]. We found that, at a $^{3}$He pressure of 100 kPa, tens of FUV photons are produced for every reacted neutron. This number is already encouraging as concerns the prospects for an absolute neutron detector based on FUV emissions *vs.* the electrical discharges that drive proportional counters, as discussed previously [11]. When mixtures of Ar, Kr or Xe are added to the $^{3}$He cell, we found much stronger FUV signals than were observed in He alone, in some cases by factors exceeding 1000. Using the spectral analysis reported below, we have identified this radiation as predominantly due to rare gas excimer ($X_2^*$) emissions,.

The experimental apparatus consists of a gas cell, photomultiplier tube (PMT) detector and gas handling system connected to a turbo-molecular/molecular-drag pump backed by an oil-free diaphragm pump. The gas cell is a stainless steel cube with 70-mm diameter metal-seal flange ports on each of the six faces. It is similar to that reported in [11] but with a modification to allow the insertion of spectral filters in front of the PMT. Sapphire, CaF$_2$ and fused silica filters were used to form a coarse spectrometer to analyze the wavelength of the emitted radiation. Neutrons enter and exit the gas cell through two fused silica windows. A 25-mm diameter thin walled magnesium tube mounted vertically in the center of the gas cell defines the neutron interaction region viewed by the detector. Both silica and magnesium are essentially transparent to neutrons, and the neutron beam is neither significantly scattered nor absorbed upon passing through these materials. Under normal operation the cell is baked for a minimum of 10 h at 70 °C while being evacuated. This removes water and other contaminants from the walls and results in a base pressure of about $3 \times 10^{-8}$ kPa. After baking and evacuating the cell to the base pressure, we introduced ultrahigh purity $^3$He into it with a gas handling system. This consists of a stainless steel manifold connected to the gas cell through a Microtorr, Model MC1-902-F filter to remove trace contaminants from the gas. Subsequent introduction of ultrahigh purity Ar, Kr, and Xe was also through Microtorr filters connected between the manifold and gas cell. Gas pressure in the evacuated cell was measured with a Pfeiffer Vacuum PKR251 gauge. An Omega DPI 705 digital pressure gauge measured the pressures of the admitted gases.

A 4-mm diameter neutron beam from the NG6-A beam line at the NIST Center for Neutron Research (NCNR) was directed into the gas cell. The neutron beam fluence was $(2.61 \pm 0.37) \times 10^5$ s$^{-1}$ cm$^{-2}$, as measured with a calibrated fission detector.

FUV radiation is detected with a solar-blind PMT (Hamamatsu R6835), operated at a bias of –2200 V and located behind a MgF$_2$ window in the gas cell. The response of the detector system is limited by the absorption edge of the MgF$_2$ and the work function of the PMT photocathode, which correspond to wavelengths of 115 nm and 190 nm respectively. The solid angle subtended by the PMT about the center of the reaction region defined by the magnesium cylinder is $(0.0373 \pm 0.0008)$ sr. No radiation produced outside of the cylinder can reach the PMT.

In our original work [11] we observed that several Lyman-$\alpha$ photons were produced for each reacted neutron when the cell was filled with a mixture of $^3$He and $^4$He. In the experiment described here, we observed a significant increase in the detected signal when Ar, Kr, and Xe were mixed with the $^3$He. In our experimental system, the $^3$He($n,tp$) reaction in the presence of these gases yields a signal of up to 1000 times greater than that which occurs in the presence of $^4$He or Ne. It is well known [4, 10, 13-15] that energetic particles traversing a noble gas will form excimers. These unstable, excited diatomic molecules radiatively dissociate with emissions in the FUV. Figure 1 reproduces the emission spectra of Ar$_2^*$, Kr$_2^*$, and Xe$_2^*$ from the work of [10]. The emission from He$_2^*$ and Ne$_2^*$ is below the absorption edge of the MgF$_2$ window and cannot be detected by our system.

To test the hypothesis that the increased signal is due to excimer formation and radiative dissociation, we used CaF$_2$, sapphire, and fused silica filters capable of discriminating among the emissions from the Ar$_2^*$, Kr$_2^*$, and Xe$_2^*$. The transmission of each filter as a function of wavelength was measured from 113 nm to 226 nm at the Far-Ultraviolet Calibration Facility [16] at NIST. The transmission and spatial uniformity of each was also measured at the normal-incidence radiometry beamline [17] at SURF III. These measurements revealed that the filters had poor spatial uniformity, but that the absorption edges were at the expected wavelengths. Figure 2 shows the responsivity of the PMT in combination with the various filters, indicating the system's utility for determining that the signal enhancement is due to excimer emission. When each filter is inserted between the MgF$_2$ widow and the PMT, the short wavelength cutoff of the detector system is shifted from 115 nm to a longer wavelength: 122 nm for CaF$_2$, 142 nm for sapphire, and 160 nm for fused silica. The Ar$_2^*$ excimer emission can be detected only with the CaF$_2$ filter in place; the other two filters are opaque to the emitted radiation. Emission from Kr$_2^*$ can be detected through CaF$_2$ and weakly through sapphire, while emission from Xe$_2^*$ can be detected through all three filters.

The experiment was performed on beamline NG6-A at the NCNR using the filter spectrometer and various noble gases. When the cell was filled with Ar, enhanced signal was seen only with the CaF$_2$ filter in place. A Kr-filled cell yielded enhanced signal with CaF$_2$ and sapphire. Significant signal gains were observed with all three filters when the cell was filled with Xe. We have modeled the expected signal enhancement using the measured filter transmissions and the excimer emission spectra from Fig. 1. As shown in Fig. 3, the observed signal enhancement in our system is consistent with our model results, providing strong evidence that the observed emissions are from noble gas excimers formed by collisions of the energetic proton and triton with noble gas atoms. It is likely that this mechanism is present when the $^3$He is mixed with $^4$He or Ne, but the excimer emission of these species is outside the spectral range of our detector.

Quantitative measurements of the photon yield were made with none of the spectral filters in place. The experimental count rates were corrected for dark current, background gamma radiation, and radiation from the direct interaction of the neutrons and the admitted noble gases. This was accomplished by taking measurements in the evacuated gas cell and with the cell filled with different pressures of the pure noble gases, but no $^3$He. In this way contributions to the signal from sources other than neutron absorption by $^3$He were removed. The data were also corrected for FUV radiation that is reflected into the photomultiplier by scattering from the wall of the magnesium cylinder. Ray tracing calculations based on a 4 mm diameter cylindrical source aligned with the neutron beam and using tabulated optical constants for MgO [18] (the surface is assumed to be oxidized rather than unreacted Mg), indicate that the ratio of scattered to directed radiation received by the PMT is between 0.14 and 0.31, depending upon the particular model of optical scattering used. The results are independent within 10% of whether a point source, line source or finite cylinder of various diameters is used.

The number of photons is calculated from the corrected count rate and the responsivity of the PMT. The PMT was calibrated at the normal-incidence radiometry beamline at SURF III as a function of wavelength from 125 nm to 210 nm. The calibration results were convolved with the emission spectrum of each excimer to determine an effective efficiency of the PMT for each of the noble gases investigated. From the detected photon flux, corrected for background, window transmission, and solid angle, we calculate the total number of photons generated. Using the known cross section of the $^3$He($n,tp$) reaction and the neutron flux, we calculate the number of neutrons reacted. The noble gases were added to a base pressure of 26 kPa of $^3$He. The photons per neutron absorbed were calculated assuming the emission spectra shown in Fig. 1. Figure 4 shows the number of photons produced for each reacted neutron. The uncertainties arise from the counting statistics, PMT calibration, and the corrections applied to the PMT signal and photon flux calculation. They are dominated by uncertainty in the scattered light correction.

At the highest noble gas pressures, the total number of photons emitted per reacted neutron was calculated from the data to be about 15,000 (Ar), 25,000 (Kr), and 33,000 (Xe). The total radiant energy produced is calculated from the mean photon energy and the photon production data. The total kinetic energy of the proton and triton are known to be 764 keV. From our data we find that the kinetic energies of the $^3$He($n,tp$) reaction products are converted into FUV radiant energy with efficiencies of 20% (Ar), 29% (Kr) or 33% (Xe). As we noted above, such high conversion efficiencies are comparable to those reported in noble gases excited by electrical discharges and particle beams. Here we have demonstrated the application of FUV excimer emissions resulting from the $^3$He($n,tp$) reaction as an efficient neutron detector.

**Acknowledgements:** We thank Steven Grantham and David Winogradoff for assistance in optical modeling, and Ping-Shine Shaw for help with optical calibrations. Mention of specific commercial products herein is for information purposes only and does not constitute endorsement by the National Institute of Standards and Technology.

**References:**

1. M.B. Chadwick, *et al.*, Nuc. Data Sheets **107**, 2931 (2006).
2. G. F. Knoll, *Radiation Detection and Measurement*, Fourth Edition, Wiley and Sons, 2010.
3. M. Salvermoser and D. E. Murnick, Appl. Phys. Lett **83**, 1932 (2003).
4. M. Salvermoser and D. E. Murnick, J. Appl. Phys. **94**, 3722 (2003).
5. J. D. Ametepe, *et al.*, J.. Appl. Phys, **85**, 7505 (1999).
6. M. Moselhy, *et al.*, Appl. Phys. Lett. **79**, 1240 (2001).
7. A. Morozov, *et al.*, J. Appl. Phys. **103**, 103301 (2008).
8. C. Duzy and J. Boness, IEEE J. Quantum Electron. **QE-16**, 640 (1980).
9. D. J. Eckstrom, *et al*, J. Appl., Phys. **64**, 1691 (1988).
10. T. E. Stewart, *et al.*, J. Opt. Soc. Am. **60**, 1290 (1970).
11. A. K. Thompson, *et al.*, J. Res. Natl. Inst. Stand. Technol. **113**, 69 (2008).
12. P. P. Hughes, *et al.*, J. Res. Natl. Inst. Stand. Technol. **114**, 185 (2009).
13. S. Kubota, T. Takahashi, and T. Doke, Phys. Rev. **165** 225 (1965).
14. P. G. Wilkinson and E. T. Byram, Appl. Opt. **4**, 581 (1965).
15. D. C. Lorents, Rad. Res. **59**, 438 (1974).
16. R. E. Vest, et al., Adv. Space Res. **37**, 283 (2006).
17. P.-S. Shaw, *et al.*, Rev. Sci. Instr. **72**, 2242 (2001).
18. D. M. Roessler and D. R. Huffman, in *Handbook of Optical Constants of Solids II*, E. D. Palik, ed. (Academic Press, Boston, 1991) pp. 919-955.

**Figure captions:**

**Figure 1**. Emission spectra of the excimers of Ar, Kr, and Xe produced by the passage of 4 MeV protons through the rare gases at a pressure of 53 kPa. Redrawn from [10].

**Figure 2**. PMT response as a function of wavelength measured at the SURF III synchrotron at NIST. Also shown are the response functions of the PMT in combination with the $CaF_2$, sapphire, and fused silica filters.

**Figure 3**. Comparison between observed and modeled relative emission through the $CaF_2$, sapphire, and fused silica filters to the PMT for 26 kPa $^3$He with added Ar, Kr, and Xe at a total pressure of 100 kPa.

**Figure 4**. Left hand scale, photons per neutron reacted for Xe, Kr, and Ar at 80 and 160 kPa with a $^3$He pressure of 26 kPa and a neutron fluence of $(2.15 \pm 0.15) \times 10^5$ $s^{-1}cm^{-2}$. The wide horizontal lines indicate the upper and lower limits of the average values taking into consideration random and systematic errors as well as uncertainties in the fraction of photons scattered into the PMT. Right hand scale, observed PMT counts from which the left hand scale is derived. The narrow horizontal lines indicate the upper and lower limits of the average values taking into consideration uncertainties in counting statistics, background, and dark current.

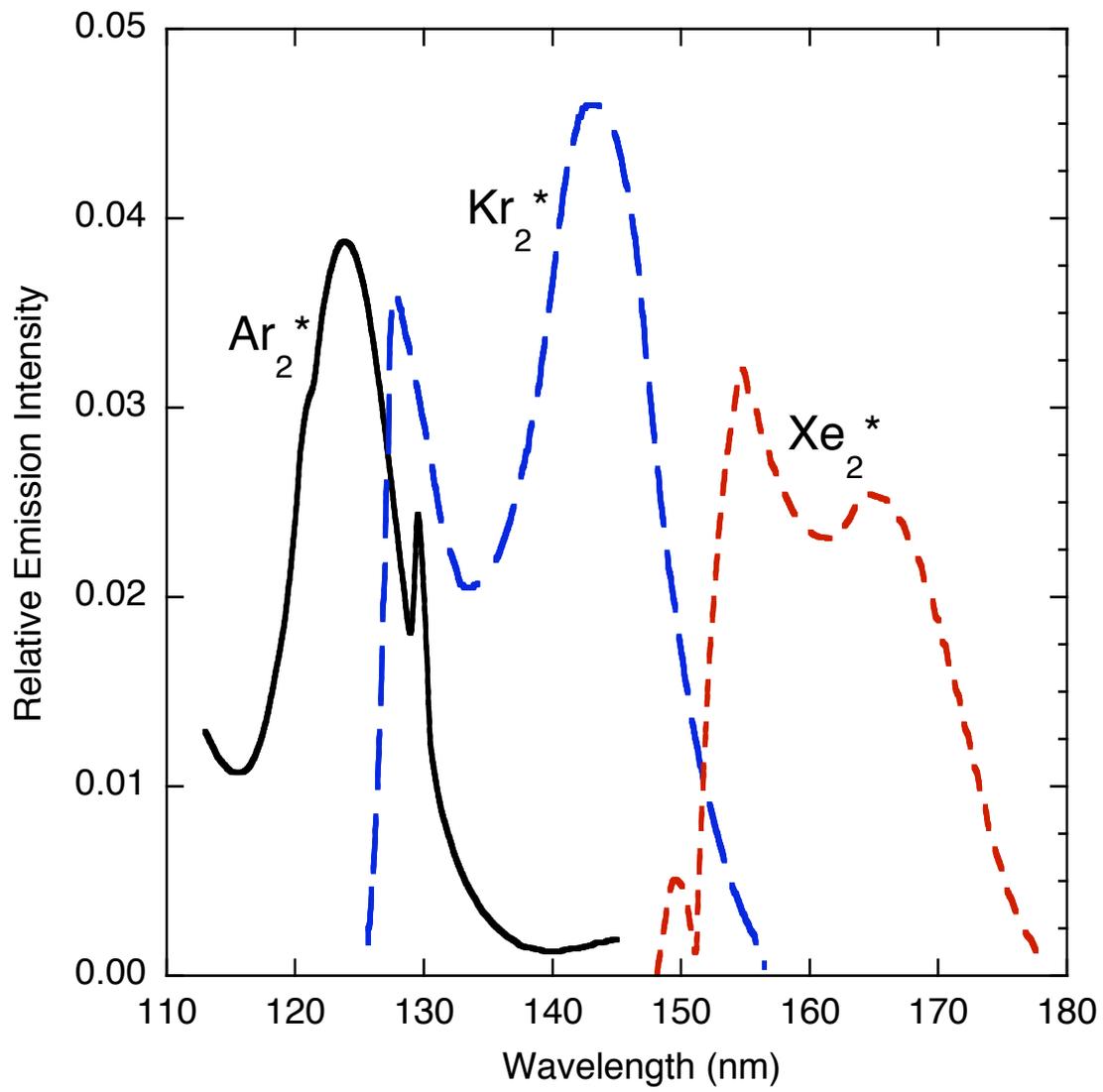

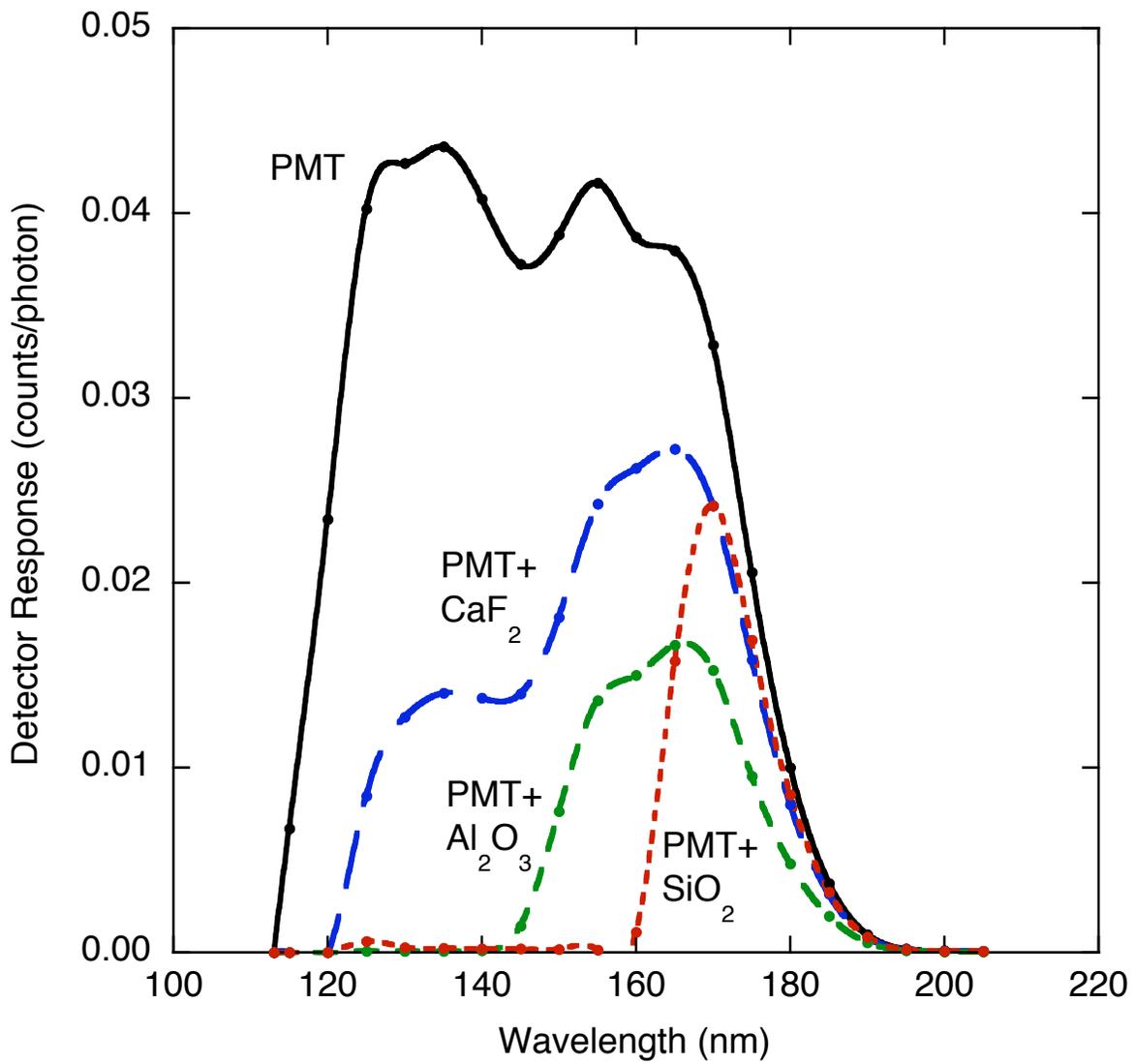

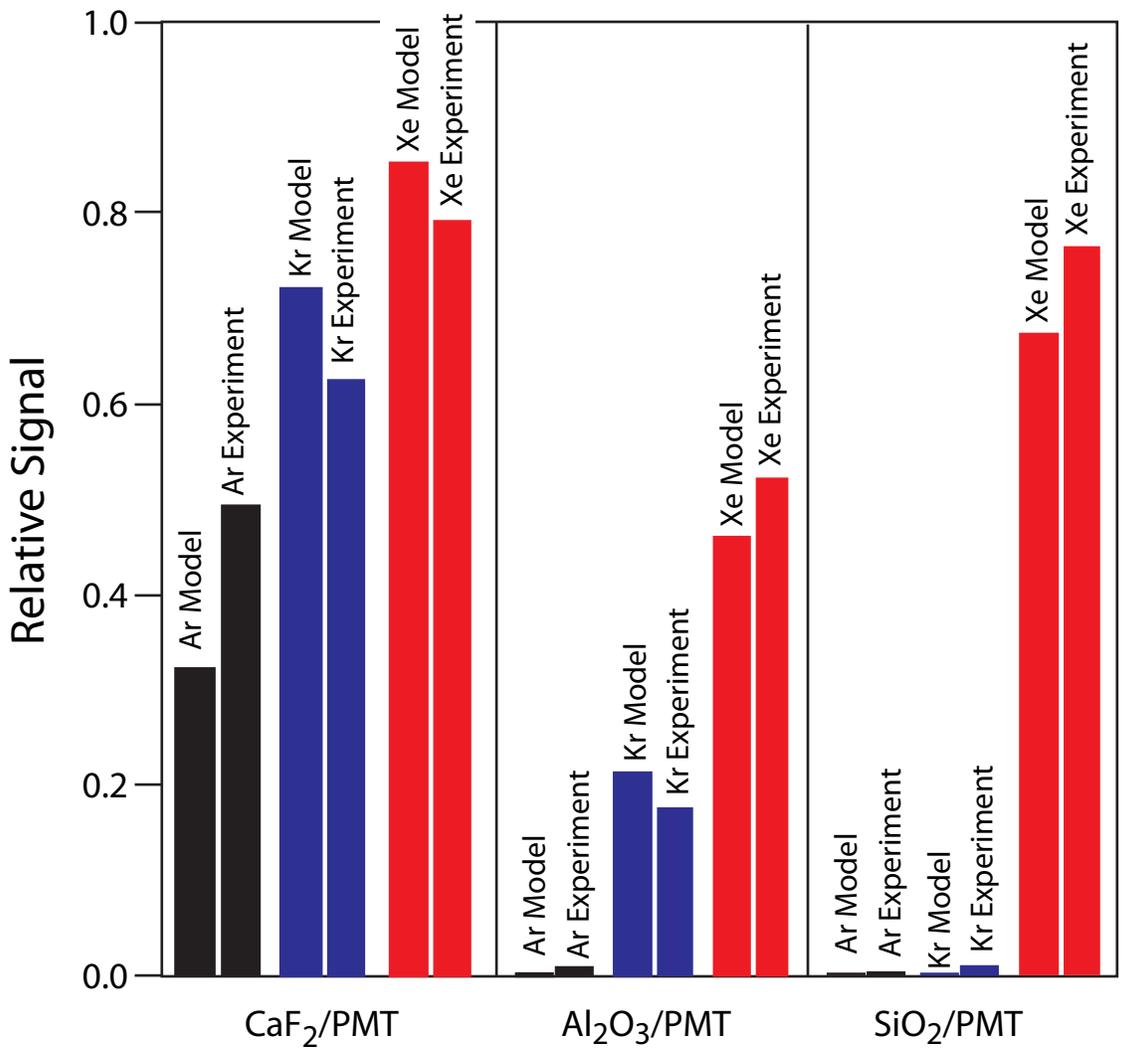

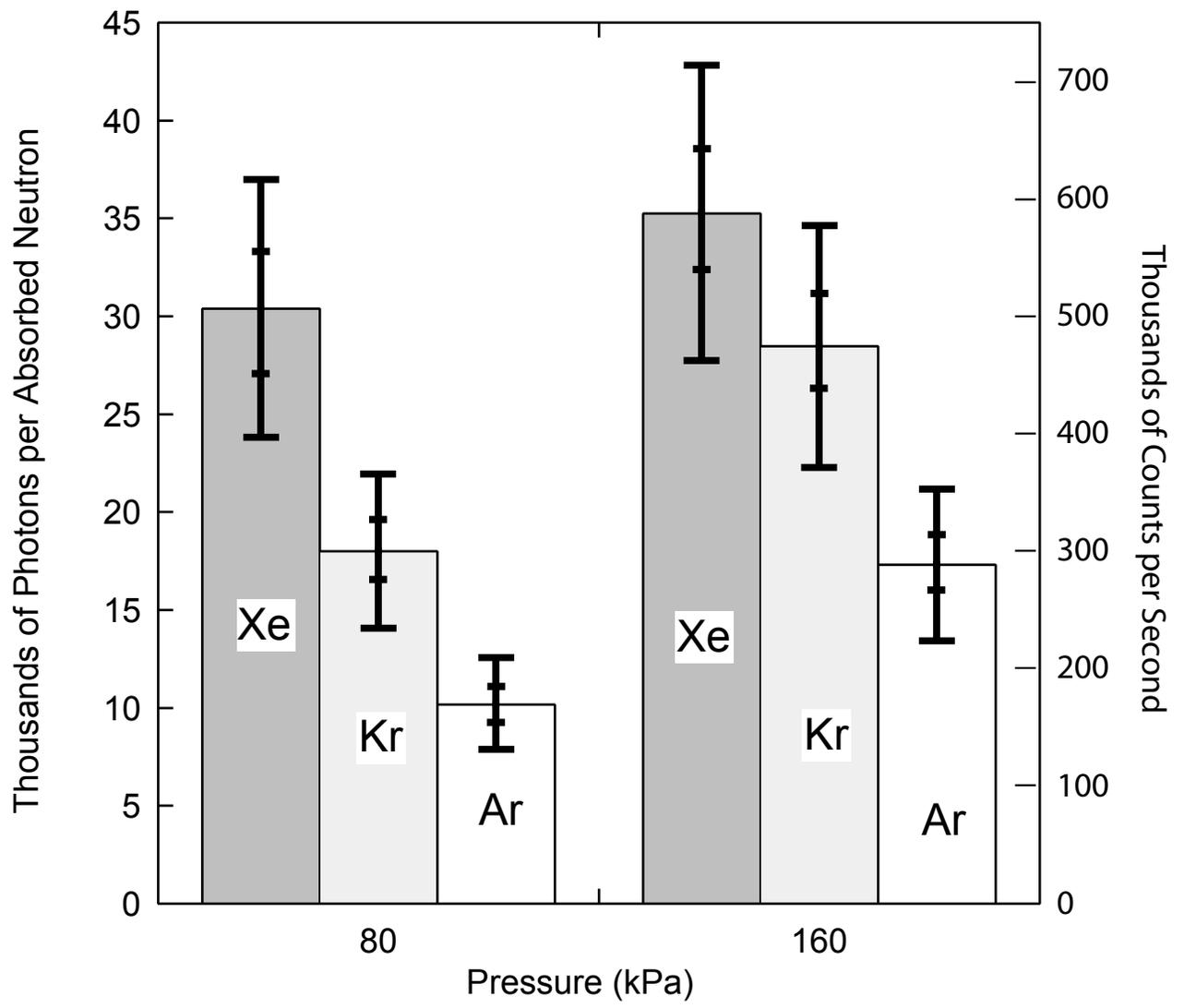